\documentclass[useAMS,usegraphicx,usenatbib]{mn2e}
\usepackage{times}
\begin{document}

\title{Bars in early- and late-type disks in COSMOS}

\author[Cameron et al.]{E.~Cameron,$^{1}$ C.~M.~Carollo,$^{1}$ P.~Oesch,$^{1}$ M.~C.~Aller,$^{1}$ T. Bschorr,$^1$ P. Cerulo,$^1$\newauthor
H.~Aussel,$^{2}$ P.~Capak,$^{3,4}$ E.~Le~Floc'h,$^{2}$ O.~Ilbert,$^{5}$ J.-P.~Kneib,$^{5}$ A.~Koekemoer,$^{6}$\newauthor  A.~Leauthaud,$^{7}$ S.~J.~Lilly,$^{1}$ R.~Massey,$^{3,8}$ H.~J.~McCracken,$^{9}$ J.~Rhodes,$^{3,10}$\newauthor  M.~Salvato,$^{3}$ D.~B.~Sanders,$^{11}$ N.~Scoville,$^{3}$ K.~Sheth,$^{3,4}$ Y.~Taniguchi$^{12}$\newauthor and D.~Thompson$^{3,13}$\\
$^{1}$Department of Physics, Swiss Federal Institute of Technology (ETH Zurich), CH-8093 Zurich, Switzerland\\
$^{2}$AIM, CNRS, Universit\'e Paris Diderot, B\^at. 709, CEA-Saclay, 91191 Gif-sur-Yvette Cedex, France\\
$^{3}$California Institute of Technology, MS 105-24, 1200 East California Boulevard, Pasadena, CA 91125, USA\\
$^{4}$Spitzer Space Center, California Institute of Technology, Pasadena, CA 91125, USA\\
$^{5}$Laboratoire d'Astrophysique de Marseille, CNRS, Universit\'e de Provence, 38
rue Fr\'ed\'eric Joliot-Curie, 13388 Marseille Cedex 13, France\\
$^{6}$STScI, 3700 San Martin Dr., Baltimore, MD 21218, USA\\
$^{7}$LBNL \& BCCP, University of California, Berkeley, CA 94720, USA\\
$^{8}$Institute for Astronomy, Royal Observatory, Edinburgh, EH9 3HJ, UK\\
$^{9}$Institut d'Astrophysique de Paris, UMR 7095, CNRS, Universit\'e Pierre et Marie Curie, 98 bis Boulevard Arago, 75014 Paris, France\\
$^{10}$Jet Propulsion Laboratory, MS 169-506, 4800 Oak Grove Drive, Pasadena, CA 91109, USA\\
$^{11}$Institute for Astronomy, 2680 Woodlawn Drive, University of Hawaii, Honolulu, HI 96822, USA\\
$^{12}$Research Center for Space and Cosmic Evolution, Ehime University, Bunkyo-cho 2-5, Matsuyama 790-8577, Japan\\\
$^{13}$LBT Observatory, University of Arizona, 933 N.\ Cherry Ave., Tucson, AZ 85721, USA}

\maketitle
\begin{abstract}
We investigate the (large-scale) bar fraction in a {\it mass-complete} sample of $M > 10^{10.5} M_\odot$ disk galaxies at $0.2 < z < 0.6$ in the COSMOS field. The  fraction of barred disks strongly depends on mass, disk morphology, and specific star formation rate (SSFR).  At intermediate stellar mass ($10^{10.5} < M < 10^{11}$ $M_\odot$) the bar fraction in early-type disks is much higher, at all redshifts, by a factor $\sim$2,  than that in late-type disks.  This trend is reversed at higher stellar mass ($M > 10^{11}$ $M_\odot$), where the fraction of bars in early-type disks becomes significantly lower, at all redshifts, than that in late-type disks.  The bar fractions for galaxies with low and high SSFRs closely follow those of the morphologically-selected early-type and late-type populations, respectively. This indicates a close correspondence  between morphology and SSFR in disk galaxies at these earlier epochs.  Interestingly, the total bar fraction in $10^{10.5} < M < 10^{11}$ $M_\odot$  disks is built up by a factor of $\sim$2 over the redshift interval explored, while for $M >10^{11}$ $M_\odot$ disks it remains roughly constant.  This indicates that, already by $z\sim0.6$, spectral and morphological transformations in the most massive disk galaxies have largely converged to the familiar Hubble sequence that we observe in the local Universe, while for intermediate mass disks this convergence is ongoing until at least $z \sim 0.2$.  Moreover, these results highlight the importance of employing mass-limited samples for quantifying the evolution of barred galaxies.  Finally, the evolution of the barred galaxy populations investigated does not depend on the large-scale environmental density (at least, on the scales which can be probed with the available photometric redshifts).
\end{abstract}


\section{Introduction}
Numerous observational studies have demonstrated that large-scale stellar bars are remarkably common amongst local disk galaxies (e.g.\ \citealt{dev91,esk00,men07,agu09}), and that a substantial population of barred disks exists out to at least redshift unity (e.g.\ \citealt{abr99,elm04,jog04,elm07,she08}).  Quantifying the fraction of barred disks as a function of redshift, and its dependence on fundamental galaxy properties, is an essential step towards understanding galaxy formation in a cosmological context.  Importantly, large-scale bars serve as signposts of massive, dynamically-cold disks \citep{ath86}, constraining their epoch of formation \citep{she08}.  Moreover, bars are key drivers of secular evolution in their host galaxies, redistributing angular momentum, enhancing nuclear star formation, and building pseudobulges (e.g.\ \citealt{lyn72,com90,kna95,car99,car97,car01,car02,car07,she00,kor04,she05,deb06,foy09,com10}).

A number of recent studies have examined the disk galaxy bar fraction for luminosity-selected samples at intermediate-to-high redshifts in the field (e.g.\ \citealt{jog04,she08}), in clusters \citep{mar09}, or both \citep{bar09}.  \citet{jog04} recovered a constant bar fraction out to redshift unity in the GEMS survey for disk galaxies at $M_V \le -19.3$ and $-$20.6 mag.  Conversely, for galaxies at $L^{\ast}_V$ ($M_V^{\ast} = -21.7$ mag at $z=0.9$) and brighter in the Cosmological Evolution Survey (COSMOS), \citet{she08} identified a build up of the barred population over the same epoch.  \citet{mar09} and \citet{bar09} have identified a minimal dependence of the bar fraction on environment (see also \citealt{men10}), except perhaps in cluster centres, where bar formation appears to be enhanced.  Both studies also recovered a rise in the bar fraction towards later morphological types, or bluer, increasingly disk-dominated systems, while \citet{she08} recovered an enhanced bar fraction in redder, increasingly bulge-dominated systems at high redshifts. However, care must be taken when comparing these results due to the different luminosity limits adopted (e.g., \citealt{mar09} and \citealt{bar09} sampled faint disks down to $M_V \le -18$ mag at $z \sim 0.165$, and $M_V \le -20$ mag  at $0.4 < z <0.8$,  respectively).  Interestingly, the highest bar fraction found by \citet{mar09} was 75$\pm$11\% for clumpy disks with distinct bulges at high luminosities ($-21 < M_V < -20$ mag).  A further complication is introduced by the various methods of disk selection employed, whether SED-fit type and visual inspection \citep{she08}, visual inspection alone \citep{bar09}, or colour, global S\'ersic index, and visual inspection \citep{mar09}.

In the light of recent evidence that galaxy evolutionary histories are tied closely to total stellar mass \citep{bun05,bal06,bol09,tas09,kov09}, the further study of evolution in mass-limited samples is a crucial step forward to undertake.  In this Letter we quantify the bar fraction at intermediate-to-high redshifts, at fixed stellar mass, using a sample of 916 morphologically-classified disk galaxies in the COSMOS field. Specifically, we study the dependence of the (large-scale) bar fraction on detailed morphological type (early- or late-type disks), specific star formation rate (SSFR), and large-scale environmental density.

The outline of this Letter is as follows.  In Section \ref{dataset} we review the COSMOS dataset employed in this work, and describe our sample selection.  In Section \ref{method} we explain our bar detection procedure and method of accounting for selection biases.  In Section \ref{results} we present our results, and in Section \ref{discussion} we discuss their implications for galaxy formation scenarios.  The construction and bar fraction analysis of a complementary sample of local Universe disk galaxies in the SDSS is described in the Appendix.  We adopt a cosmological model with $\Omega_\Lambda = 0.75$, $\Omega_M = 0.25$, and $h=0.7$, and all magnitudes are quoted in the AB system throughout.

\section{Dataset and ancillary measurements}\label{dataset}

We use the COSMOS dataset \citep{sco07}, consisting of ground-based and space-based, multi-wavelength imaging of an $\sim$2 deg$^2$ equatorial field.  In particular, as the starting point for our sample definition,  we adopt the $I$-band source catalogue of \citet{lea07} extracted from the COSMOS ACS $I$-band imaging frames \citep{koe07}, consisting of 156,748 sources limited down to  $I_\mathrm{AB} = 26.6$  mag.  Careful artifact and star removal was performed to improve the robustness of this catalogue, and computation of photometric redshifts was attempted for 111,141 of these sources flagged as galaxies (i.e., removing stars and junk).   

Eleven photometric bands covering the wavelength range between the COSMOS CFHT $u$ and the Spitzer $4.5$ $\micron$ band (see \citealt{cap07,san07}) were used to derive these photometric redshifts using our own  {\it Zurich Extragalactic Bayesian Redshift Analyzer}  (ZEBRA\footnote{The code is publicly available under the following URL: \texttt{www.exp-astro.phys.ethz.ch/ZEBRA}}) code \citep{fel06}. To this end, we used galaxies with secure spectroscopic redshifts from the zCOSMOS survey \citep{lil07,lil09} and modified with ZEBRA our set of empirical templates \citep{col80,kin96} to account for systematic, template-dependent mismatches with the photometric spectral energy distributions (SEDs) of the sample galaxies. An acceptable template match was able to be identified for all but 1695 of the 111,141 input sources (a 1.5\% failure rate).  The resulting photometric redshifts have an uncertainty of $\Delta(z)/(1+z)\sim0.023(1+z)$ down to $I<22.5$ mag, as directly tested with available spectroscopic redshifts from zCOSMOS. We estimated the photometric redshift accuracy at magnitudes fainter than  $I=22.5$ mag by dimming the photometry of zCOSMOS galaxies down to $I=24$ mag; this results in an uncertainty of $0.039(1+z)$. As a cross-check, we also used the photometric redshifts of \citet{ilb09}, which became available in the meantime. These are based on a photometric catalog of the COSMOS survey with 30 photometric bands including also narrow- and intermediate-band filters, and the effects of both dust extinction and emission lines on template fitting were accounted for carefully during their computation.  Based on the comparison with the spectroscopic redshifts of the $I<22.5$ mag zCOSMOS sample, the dispersion of the \citet{ilb09} photo-$z$ catalog is $\sim$$0.007(1+z)$ (and $\sim$$0.012(1+z)$ down to $I=24$ mag; \citealt{ilb09}).  Owing to the use of a much larger number of photometric passbands, the \citet{ilb09} redshifts have a higher accuracy than our ZEBRA estimates. The advantage of the latter, however, is that they were derived by ourselves, and we are thus fully aware of their caveats and limitations. We therefore take the conservative strategy  to use both versions of the COSMOS photo-$z$s in our analysis, as a means to identify any systematic uncertainty associated with the photometric redshift estimates. 

In order to estimate stellar masses from the photometric data, we used a newly-developed, non-public extension of our ZEBRA code, ZEBRA+, which estimates  galaxy physical parameters, such as stellar masses,  by fitting synthetic stellar population model SEDs to the galaxy broadband photometry (cf.\ \citealt{fon04,poz07,max09}). As in ZEBRA's Maximum Likelihood mode, the best fit SED and its normalisation, are computed by minimising the $\chi^2$ between the observed and template fluxes. Additionally, ZEBRA+ includes reddening of the input SEDs with a variety of dust laws, when required.  Using the zCOSMOS 10k sample \citet{bol09} have demonstrated a level of uncertainty of $\sigma_{\log M} \approx 0.20$ arising from the choices of SED template and dust extinction law for stellar masses derived via SED-fitting to the COSMOS photometric data.  This is consistent with the scatter of $\sigma_{\log M} \approx 0.27$ between our ZEBRA+ stellar mass estimates and those from the publicly-available catalog of \citet{pan09} for galaxies well-matched in both position (centroid difference less than $2 \times 10^{-5}$ arcsec) and photometric redshift ($|z_\mathrm{ZEBRA}-z_\mathrm{Pannella}| < 0.02 (1+z_\mathrm{ZEBRA})$).  Details for ZEBRA+, including a further investigation into the relevant uncertainties, are given in Oesch et al.\ (2010, in prep.; see also \citealt{oes09}).  An acceptable synthetic template match for computation of stellar masses was able to be identified for all but 7033 of the 109,446 galaxies with estimated photometric redshifts (a 6\% failure rate).  We stress that our results remain unchanged when using either of the two photo-$z$ catalogs; furthermore, the resulting uncertainties in the stellar mass and star formation rate estimates are negligible.  A total of 32,840 galaxies were identified with photometric redshifts in the range $0.2 < z < 0.6$ examined in this study, of which 3519 are above the adopted mass limit of $M > 10^{10.5}$ $M_\odot$.

\begin{figure}
\includegraphics[width=8.4cm]{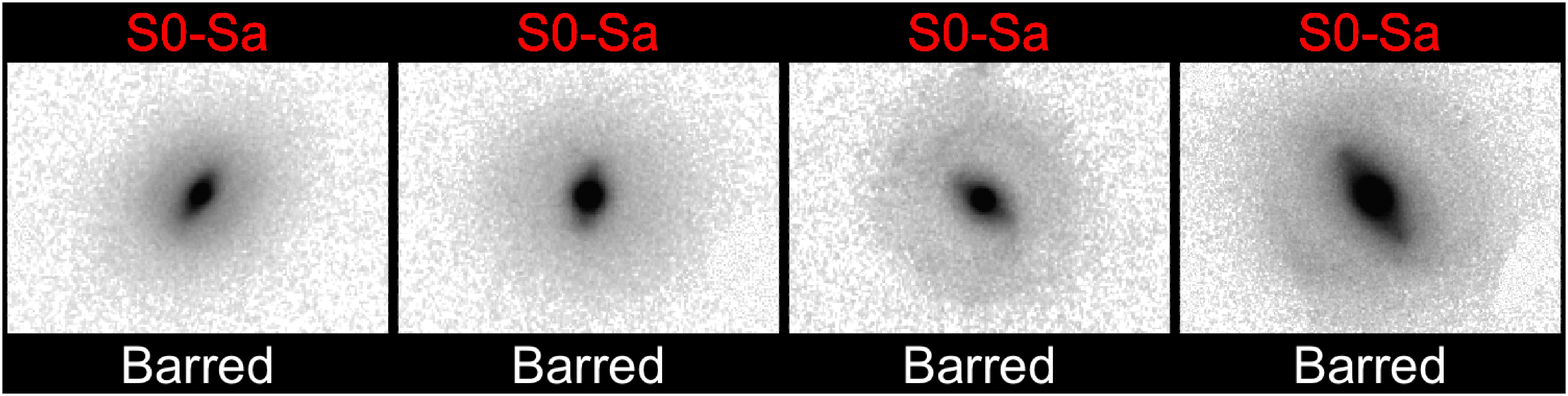}
\includegraphics[width=8.4cm]{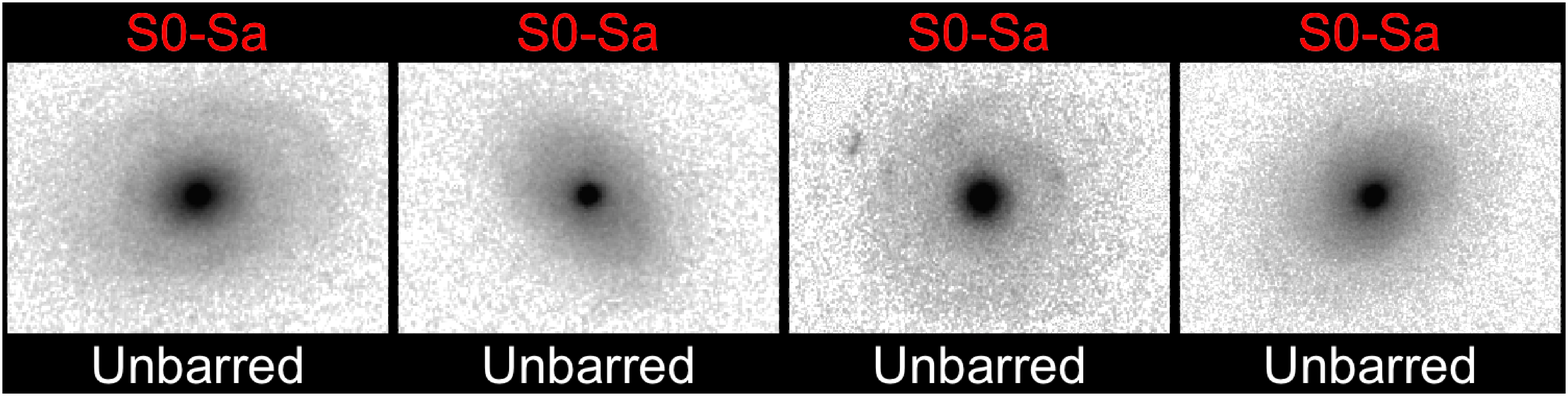}
\includegraphics[width=8.4cm]{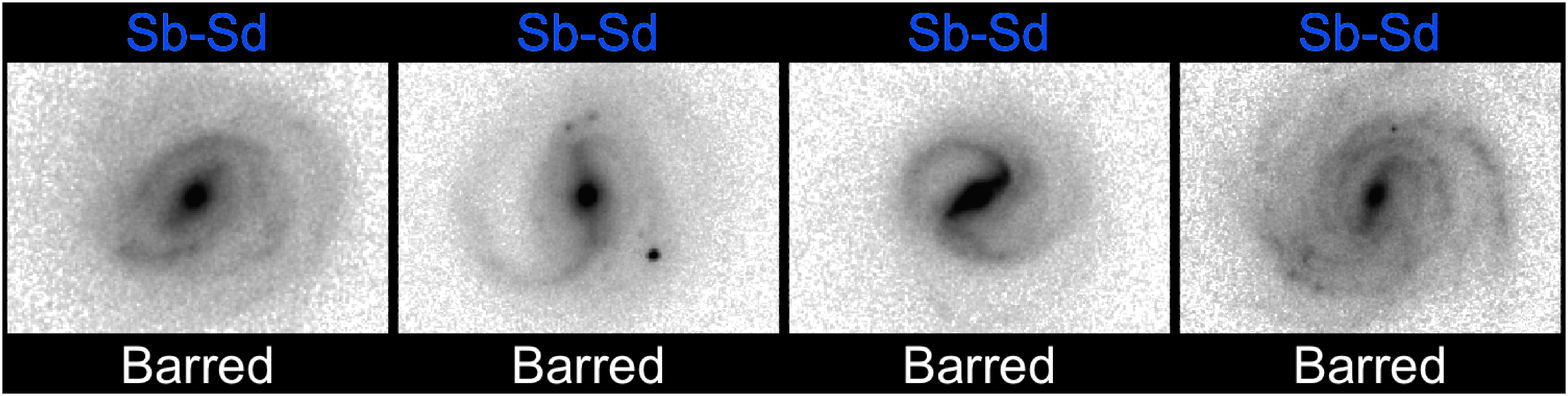}
\includegraphics[width=8.4cm]{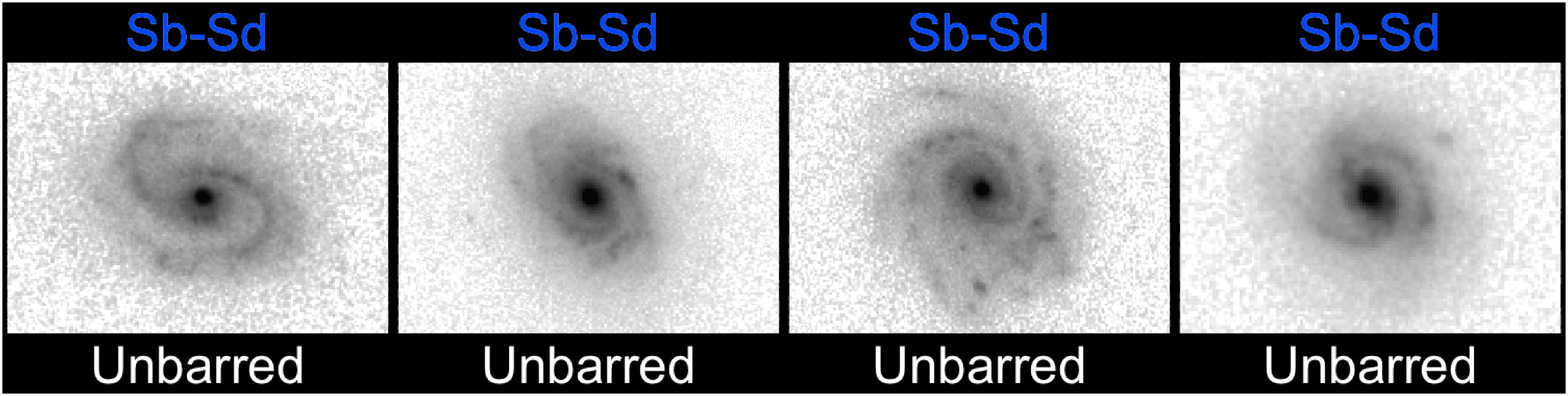}
\caption{Example (HST ACS $I$-band) postage stamp images of both early-type (\textit{top two rows}) and late-type (\textit{bottom two rows}), barred and unbarred galaxies from our sample of 916 COSMOS disks at $0.2 < z < 0.6$, $M > 10^{10.5}$ $M_\odot$, and $e < 0.3$.  The width of each of these postage stamps was adjusted within a range of 3.3 to 6.3 arcsec in angular size to best emphasise the morphological details of each system.}
\label{postage}
\end{figure}

Star formation rates were computed based upon each galaxy's total IR ($L_\mathrm{IR}$) luminosity  (derived from the MIPS detections at $24$ $\mu$m, \citealt{lef09},  down to a flux limit of $0.15$ mJy, using the templates of \citealt{rie09}) and total UV  ($L_\mathrm{UV}$) luminosity (estimated by interpolating the observed SEDs to 2800$\AA$).  Specifically, we adopt the relation $\textrm{SFR}=9.8\times 10^{-11}(L_\mathrm{IR}+2.2L_\mathrm{UV})$ $M_\odot$ yr$^{-1}$ from \citet{bel05}.  The total uncertainties in our SFR estimates are $\sim$0.4 dex, given the $\sim$0.3 dex intrinsic scatter about this relation \citep{bel05} and the $\sim$0.3 dex uncertainty on the estimation of UV luminosities from our SED template fits.  Environmental densities are sourced from Scoville et al.'s (2007) catalog of (large-scale) surface densities for COSMOS galaxies in fixed redshift intervals.

Galaxy morphologies were obtained with an upgraded version of  the \textit{Zurich Estimator of Structural Types} (ZEST; \citealt{sca07}), known as ZEST+ (Carollo et al.\ 2010, in prep).  Relative to its predecessor, ZEST+ includes additional measurements of non-parametric morphological indices for characterising both structure and substructure.  Moreover, ZEST+ offers a Support Vector Machine (SVM) classification (adopted herein), as well as a Principal Component Analysis (PCA) scheme for consistency with the earlier ZEST.  The ZEST+ SVM classifies galaxies in seven morphological types located in specific regions of the 6-dimensional space of concentration, asymmetry, clumpiness, M$_{20}$, Gini coefficient, and ellipticity.  The different types were visually inspected to ensure a broad equivalence with the following Hubble types: E, S0-Sa, Sb, Sc, Sd, and Irr (with the Irr types divided into two classes, `concentrated' and `non-concentrated', according to the value of the concentration index).  As a result of visual inspection of the 3519 galaxies in our sample at $0.2 < z  < 0.6$ and $M > 10^{10.5}$ $M_\odot$, a total of 36 Irr, 14 S0-Sa, and 33 spiral (Sa-Sb to Sc-Sd) systems were identified as `catastrophic' misclassifications and were manually reassigned to their respective types.  The two most frequent causes of such `catastropic' misclassifications were the presence of bright, overlapping neighbours and/or strong dust lanes (see \citealt{sar10} for a detailed study of the impact of dust on disks in COSMOS), which can bias recovery of the quantitative morphological indices used as input to the SVM analysis module.  Regarding the reliability of ZEST+ output in general, calibrations performed by \citet{sca07} using the original ZEST code to classify real COSMOS galaxies show that down to $I = 24$ mag the misclassification rate is at most $\sim$30\% for ellipticals incorrectly identified as early- and intermediate-type disks in the lowest S/N bin.   As ZEST+ features substantially improved algorithms for computation of the relevant quantitative morphological indices relative to the original ZEST, we are confident that the ZEST+ morphological classifications are even more reliable than those of ZEST.

The end result of our classification process was a final sample of 2820 disk galaxies (morphological type between S0-Sa and Sc-Sd), of which 916 have ellipticities ($e$) less than 0.3 (equivalent to disk inclination less than 45 deg).  Example (HST ACS $I$-band) postage stamp images of both early-type and late-type, barred and unbarred systems from our final sample of 916 COSMOS disks at at $0.2 < z < 0.6$, $M > 10^{10.5}$ $M_\odot$, and $e < 0.3$ are presented in Fig.\ \ref{postage}.

\section{Method}\label{method}

\subsection{Bar Detection}\label{detection}
Barred galaxies in our sample were identified as follows.  First, the \texttt{ellipse} package in \texttt{IRAF} was used to recover the elliptical isophote profile of each galaxy from its ACS $I$-band image.  Specifically, \texttt{ellipse} was run repeatedly for each galaxy, varying the initial guesses of object position, semi-major axis, ellipticity, and position angle until a robust fit was recovered down to the level of sky noise (cf.\ \citealt{reg97,lai02,men07}).  If a robust fit could not be identified with a single combination of initial parameter guesses ($<$5\% of cases), the profile was constructed piece-wise from the robust sections of two or more fitting attempts.  Pixels within 2.5 times the Kron radius of any nearby neighbours were excluded from the fitting process.

All galaxies with profiles displaying a monotonic increase in ellipticity of $e_{\mathrm{min}} \ge 0.4$ with a subsequent drop of at least $\Delta e \ge 0.1$ were flagged as candidate barred systems.  The threshold of $e_\mathrm{min} \ge 0.4$ was chosen to replicate the selection of strong galaxy bars made by \citet{jog04} and \citet{she08}.  However, unlike these other studies, we did not reject candidate barred systems based on a further criterion of isophote position angle change at the bar end.  Rather, the final selection of barred galaxies was performed via visual inspection of the candidates, thereby maximising the completeness of our bar sample (i.e., allowing for detection of those galaxies in which, by chance, the uncorrelated bar and disk position angles are closely aligned; cf.\ Men\'endez-Delmestre et al.\ 2007).  In total, 213 (67\%) of our 320 candidate barred systems (23\% of the entire disk galaxy sample) were thereby confirmed as barred.  A further 23 barred candidates that were ultimately rejected were noted as `ambiguous' cases (e.g.\ the `bar' morphology was perhaps more consistent with that of a lens or truncated inner disk).  Example (HST ACS $I$-band) postage stamp images of both early-type and late-type, barred and unbarred galaxies from our final sample of 916 COSMOS disks at at $0.2 < z < 0.6$, $M > 10^{10.5}$ $M_\odot$, and $e < 0.3$ are presented in Fig.\ \ref{postage}.

\subsection{Accounting for Selection Effects}\label{selection}

Completeness limits on bar detection given the isophotal ellipticity criteria described above were estimated via artificial galaxy simulations.  Both early-type and late-type barred galaxy models were constructed from three components: a S\'ersic bulge, a (truncated) S\'ersic bar, and an exponential disk.  The properties of each model were chosen to be consistent with the observed properties of barred galaxies in the local universe \citep{elm96,lau07,gad08,wei09}, and are listed in Table \ref{modelstats}.  The use of three different early type models was necessary to replicate the variety of structural types observed amongst local S0 and Sa systems (e.g.\ \citealt{bal03}).  Importantly, the models adopted span a wide range of bulge S\'ersic indices ($1 \le n \le 4$) and bulge-to-total flux ratios ($0.1 \le \mathrm{B/T} \le 0.6$), and thus encompass the entire parameter space observed for massive local disk galaxies.

For all models we adopt a bar ellipticity of $e=0.6$ as this represents a lower limit on the intrinsic bar ellipticities typically recovered for massive disk galaxies in 2D bulge-bar-disk structural decomposition studies (e.g. Gadotti 2008, MNRAS).  The lower the intrinsic ellipticity of the bar, the lower the mean surface brightness (at a given Bar/T), so it follows that that low ellipticity bars will be most sensitive to surface brightness-dependent detection biases.  Moreover, the lower the intrinsic bar ellipticity, the more sensitive will the system be to size-dependent detection biases caused by the `rounding' of bar isophotes by the PSF.  Hence, by using a `lower limit' bar ellipticity of $e=0.6$ in all models we best expose the limits of our detection routine, and thereby identify safely conservative selection limits.

\begin{table}
\caption{Key Model Parameters for Artificial Galaxy Simulations of Bar Detection Completeness}
\label{modelstats}
\begin{tabular}{llcccccc}
\hline \hline
\# & Type & B/T & Bar/T & $R_{e}/h$$^{a}$ & $n$$^{a}$ & $n_\mathrm{bar}$$^{a}$  & $L_\mathrm{bar}/h$$^a$\\
\hline
1 & Early & 0.4 & 0.15 & 0.2 & 4.0 & 0.7 & 1.15\\
2 & Early & 0.2 & 0.15 & 0.2 & 2.5 & 0.7 & 1.15\\
3 & Early & 0.6 & 0.15 & 0.2 & 4.0 & 0.7 & 1.15\\
4 & Late & 0.1 & 0.10 & 0.1 & 1.0 & 0.9 & 0.90\\
\hline
\end{tabular}
$^a$ These symbols refer to the effective radius ($R_e$), S\'ersic index ($n$), and truncation length ($L$) of the generalised 2D S\'ersic profile (cf.\ Aguerri et al.\ 2009; Graham \& Driver 2005) employed for the model bulge and bar components, while $h$ is the scalelength of the underlying exponential disk.  In all cases we adopt ellipticities ($e$) of $e=0$ and $e_\mathrm{bar}=0.6$, and shape parameters ($c$) of $c=2$ (elliptical isophotes) and $c_\mathrm{bar}= 2.4$ (boxy isophotes), for the bulge and bar respectively.  Note that in each artificial galaxy image generated the bar length is scaled relative to the projection of the disk.
\end{table}

\begin{figure*}
\includegraphics[width=15cm]{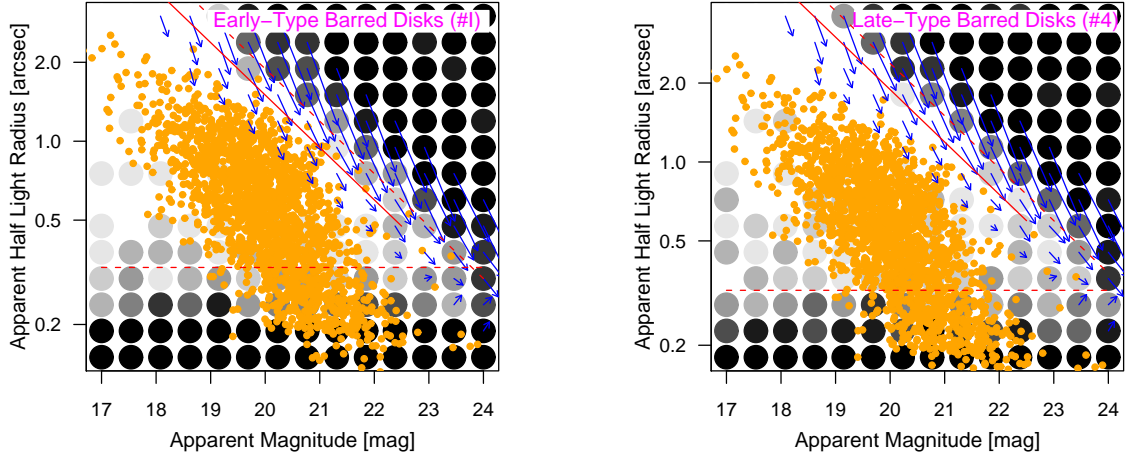}
\caption{The recoverability of bars in the COSMOS ACS $I$-band imaging for early-type (\textit{left}) and late-type (\textit{right}) model disks at low inclination ($i < 45$ deg) as derived via artificial galaxy simulations.  The detection completeness of input bars identified via our isophotal ellipse-fitting procedure in each bin of intrinsic (i.e., input) apparent magnitude and half light radius is indicated by the large grey-shaded circles on a scale from white representing 100\% completeness to black representing 0\% completeness.  The effective limits on (intrinsic) apparent size and surface brightness for high detection completeness are highlighted by dashed, red lines.  Blue arrows indicate the mean errors on the measurement of (Kron-style) galaxy magnitudes and sizes towards low surface brightnesses, and red, solid lines mark boundaries within which galaxies are no longer scattered in (due to measurement errors) from regions of low detection completeness.  Finally, the orange datapoints reveal the observed distribution of ZEST+ classified disk galaxies with $M > 10^{10.5}$ $M_\odot$  and $i < 45$ deg at $0.2 < z < 0.6$ in the COSMOS field.  Note that only the results for simulations corresponding to our models \#1 and \#4 (see Table \ref{modelstats}) are displayed here; the results for model \#2 were nearly identical to those for model \#1, while almost no bars were recovered for model \#3 (see the discussion in Section \ref{selection}).}
\label{barsims}
\end{figure*}

Artificial galaxy images for each type of model were generated and convolved with a Gaussian PSF of $0.12$ arcsec (4 pixels) FWHM, then degraded to the noise level of the COSMOS ACS $I$-band imaging.  A total of 1960 images were constructed for each model in a grid of total galaxy magnitudes and disk scalelengths spanning $I=17$ to 24 mag and $h=0.15$ to 3 arcsec (5 to 100 pixels), respectively.  Ten instances of each model were generated at each grid point with random disk inclinations ranging 0 to 45 deg, and random component position angles (with the bar size scaled relative to the projection of the disk).  These images were run through our candidate bar detection pipeline, and the results are plotted as a function of total galaxy magnitude and total half light radius (more easily comparable against observations than disk scalelength) in Fig.\ \ref{barsims}.

These simulations indicate a high bar detection completeness for most galaxies in our sample, supposing a mix of early-type and late-type disks equivalent in structure to our models \#1, \#2, and \#4 (see Table \ref{modelstats}).  Although simulations with early-type model \#3 revealed that our detection procedure is unlikely to discover bars in any severely bulge-dominated systems (B/T $\ga$ 0.6), high-resolution imaging studies indicate a negligible large-scale bar fraction for such galaxies in any case (e.g.\ \citealt{lau07,wei09}).  Comparing the distribution of ZEST+ classified disks with $M > 10^{10.5}$ $M_\odot$ and $i < 45$ deg at $0.2 < z < 0.6$ in the COSMOS field against the model completeness limits at low surface brightness (for models \#1, \#2, and \#4, see Fig.\ \ref{barsims}) reveals that bar detection in our sample is not surface brightness limited.  In particular, there are very few galaxies with observed magnitudes and half light radii approaching these limits, even after accounting for the systematic biases at low signal-to-noise inherent in the Kron-style measurements employed here (cf.\ \citealt{cam09}).  However, the limit for bar detection in galaxies with small apparent half light radii ($R_e \sim 0.33$ arcsec) is significant and must be considered in our analysis.

To ensure a uniform detection completeness of barred disks at all redshifts we exclude from our subsequent analysis all galaxies smaller than the physical size represented by this apparent size limit at the highest redshifts explored here (i.e., $R_{e,\mathrm{lim}}=2.2$ kpc at $z=0.6$).  This cut removes 283 ($\sim$31\%) of our inital sample of 916 COSMOS disks, including 30 bars confidently detected in (primarily early-type) galaxies with $R_e \sim 1.5$$-$$2.2$ kpc at $z \la 0.45$.  Hence, we can confirm that bars are observed to exist in compact disks at these intermediate redshifts, but note that the evolution of this population cannot be robustly constrained with the available data.  Therefore, we emphasise that the results presented in this study concern intermediate-to-large disks ($R_e > 2.2$ kpc) only.

\section{Results}\label{results}

\begin{figure*}
\center
\includegraphics[width=17.5cm]{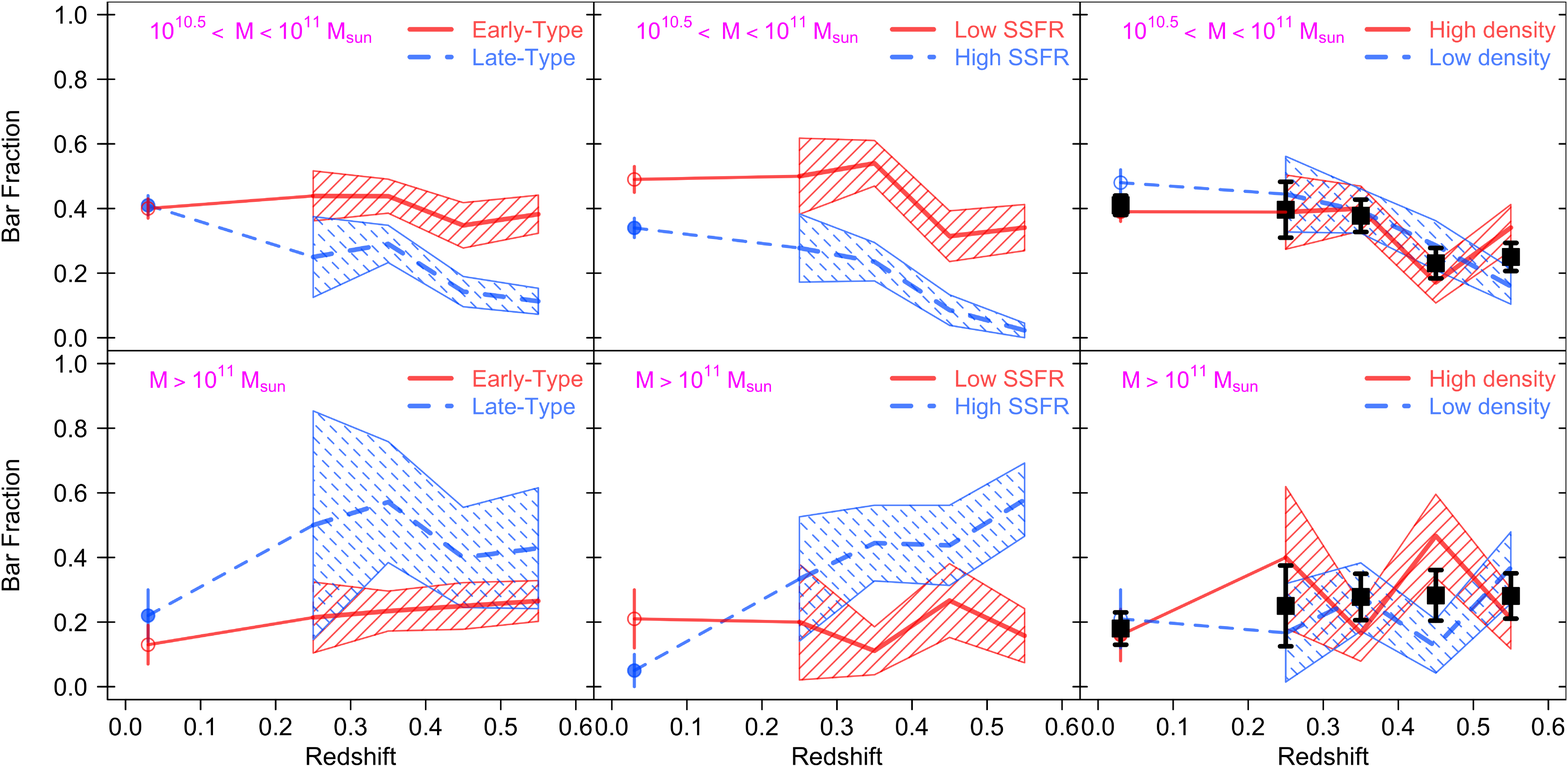}
\caption{The redshift evolution of the bar fractions of ZEST+ classified disk galaxies with $M > 10^{10.5}$ $M_\odot$, $i < 45$ deg, and $R_{e} > 2.2$ kpc at $0.2 < z < 0.6$ in the COSMOS field subdivided by stellar mass, morphological type, SSFR, and environmental density.  Galaxies at intermediate stellar mass ($10^{10.5} < M < 10^{11}$ $M_\odot$) are shown on the top row, while those at high stellar mass ($M > 10^{11}$ $M_\odot$) are shown on the bottom row.  The left column displays the bar fraction evolution by ZEST+ morphological type for early-type and late-type disks.  The middle and right columns display the bar fraction evolution by SSFR and (large-scale) environmental density respectively for the lowest and highest thirds of the population in each mass and redshift bin.  In each panel the thick red and blue lines indicate the bar fractions of the relevant populations, and the corresponding uncertainties are revealed by the cross-hatched areas.  The black squares and error bars overlaid on the plots by density indicate the total disk galaxy bar fraction in each redshift bin. Note the lack of evolution at $M>10^{11} M_\odot$, and the net evolution below this mass scale.  The local Universe bar fractions marked at $z=0.03$ indicate the values measured in our low redshift ($0.02 < z < 0.04$) SDSS comparison sample selected and analysed as described in the Appendix.}
\label{mass}
\end{figure*}

\begin{table*}
\caption{Bar Fractions by Mass, Morphology, SSFR, and Environmental Density in COSMOS at $0.2 < z < 0.6$}
\label{cosmosfracs}
\begin{tabular}{llrrr}
\hline \hline
Redshift & Stellar Mass$^a$ & Morphological Type & Specific Star Formation Rate$^a$ & Environmental Density$^a$\\
\hline
$0.2 < z < 0.3$ & Intermediate & Early-Type : $f_\mathrm{bar} = 0.44 \pm 0.08 $ & Low SSFR :  $f_\mathrm{bar} = 0.50 \pm 0.11 $ & High density : $f_\mathrm{bar} =  0.44  \pm 0.11 $\\
$0.2 < z < 0.3$ & Intermediate & Late-Type : $f_\mathrm{bar} = 0.25 \pm 0.13 $ & High SSFR :  $f_\mathrm{bar} = 0.27 \pm 0.11 $ & Low density : $f_\mathrm{bar} =  0.38 \pm 0.11 $\\
$0.2 < z < 0.3$ & High & Early-Type : $f_\mathrm{bar} = 0.21 \pm 0.11 $ & Low SSFR :  $f_\mathrm{bar} = 0.20 \pm 0.18 $ & High density : $f_\mathrm{bar} =  0.16  \pm 0.15 $\\
$0.2 < z < 0.3$ & High & Late-Type : $f_\mathrm{bar} = 0.50 \pm 0.35 $ & High SSFR :  $f_\mathrm{bar} = 0.33 \pm 0.19 $ & Low density : $f_\mathrm{bar} =  0.40  \pm 0.21 $\\
\hline
$0.3 < z < 0.4$ & Intermediate & Early-Type : $f_\mathrm{bar} = 0.44 \pm 0.05 $ & Low SSFR :  $f_\mathrm{bar} = 0.54 \pm 0.07 $ & High density : $f_\mathrm{bar} =  0.39  \pm 0.07 $\\
$0.3 < z < 0.4$ & Intermediate & Late-Type : $f_\mathrm{bar} = 0.29 \pm 0.06 $ & High SSFR :  $f_\mathrm{bar} = 0.24 \pm 0.06 $ & Low density : $f_\mathrm{bar} =  0.40  \pm 0.07 $\\
$0.3 < z < 0.4$ & High & Early-Type : $f_\mathrm{bar} = 0.23 \pm 0.06 $ & Low SSFR :  $f_\mathrm{bar} = 0.11 \pm 0.07 $ & High density : $f_\mathrm{bar} =  0.28  \pm 0.10 $\\
$0.3 < z < 0.4$ & High & Late-Type : $f_\mathrm{bar} = 0.57 \pm 0.18 $ & High SSFR :  $f_\mathrm{bar} = 0.44 \pm 0.12 $ & Low density : $f_\mathrm{bar} =  0.17  \pm 0.09 $\\
\hline
$0.4 < z < 0.5$ & Intermediate & Early-Type : $f_\mathrm{bar} = 0.35 \pm 0.07 $ & Low SSFR :  $f_\mathrm{bar} = 0.31 \pm 0.07 $ & High density : $f_\mathrm{bar} =  0.29  \pm 0.07 $\\
$0.4 < z < 0.5$ & Intermediate & Late-Type : $f_\mathrm{bar} = 0.14 \pm 0.05 $ & High SSFR :  $f_\mathrm{bar} = 0.08 \pm 0.05 $ & Low density : $f_\mathrm{bar} =  0.17  \pm 0.06 $\\
$0.4 < z < 0.5$ & High & Early-Type : $f_\mathrm{bar} = 0.25 \pm 0.07 $ & Low SSFR :  $f_\mathrm{bar} = 0.27 \pm 0.11 $ & High density : $f_\mathrm{bar} =  0.13  \pm 0.08 $\\
$0.4 < z < 0.5$ & High & Late-Type : $f_\mathrm{bar} = 0.40 \pm 0.15 $ & High SSFR :  $f_\mathrm{bar} = 0.44 \pm 0.12 $ & Low density : $f_\mathrm{bar} =  0.46  \pm 0.13 $\\
\hline
$0.5 < z < 0.6$ & Intermediate & Early-Type : $f_\mathrm{bar} = 0.38 \pm 0.06 $ & Low SSFR :  $f_\mathrm{bar} = 0.34 \pm 0.07 $ & High density : $f_\mathrm{bar} =  0.16  \pm 0.05 $\\
$0.5 < z < 0.6$ & Intermediate & Late-Type : $f_\mathrm{bar} = 0.11 \pm 0.04 $ & High SSFR :  $f_\mathrm{bar} = 0.03 \pm 0.03 $ & Low density : $f_\mathrm{bar} =  0.34  \pm 0.07 $\\
$0.5 < z < 0.6$ & High & Early-Type : $f_\mathrm{bar} = 0.26 \pm 0.06 $ & Low SSFR :  $f_\mathrm{bar} = 0.16 \pm 0.08 $ & High density : $f_\mathrm{bar} =  0.37  \pm 0.11 $\\
$0.5 < z < 0.6$ & High & Late-Type : $f_\mathrm{bar} = 0.43 \pm 0.19 $ & High SSFR :  $f_\mathrm{bar} = 0.58 \pm 0.11 $ & Low density : $f_\mathrm{bar} =  0.21  \pm 0.09 $\\
\hline
\end{tabular}
\flushleft{$^a$Here we define intermediate stellar mass as $10^{10.5} < M < 10^{11}$ $M_\odot$ and high stellar mass as $M > 10^{11}$ $M_\odot$, and we identify the low and high SSFR (or environmental density) populations as the lower and upper thirds of the SSFR (or environmental density) distribution in each mass and redshift bin (as explained in Section \ref{results}).  The COSMOS disk sample examined is also restricted to systems with $e < 0.3$ and $R_e > 2.2$ kpc as described in Section \ref{selection}.}\\
\end{table*}

In Fig.\ \ref{mass} we present the (strong) bar fraction of ZEST+ classified disk galaxies with $M > 10^{10.5}$ $M_\odot$, $i < 45$ deg, and $R_{e} > 2.2$ kpc in the COSMOS field in redshift bins of $0.2 < z < 0.3$, $0.3 < z < 0.4$, $0.4 < z < 0.5$, and $0.5 < z < 0.6$.  Two key stellar mass regimes are explored: intermediate mass ($10^{10.5} < M < 10^{11}$ $M_\odot$), and high mass ($M > 10^{11}$ $M_\odot$).  Within each mass regime we further explore the dependence of bar fraction on (ZEST+ classified) morphological type (early-type vs.\ late-type), specific star formation rate, and (large-scale) environmental density.  For the purposes of exploring the latter two redshift dependencies we identify the lower and upper thirds of the population in each parameter in each mass and redshift bin.  We also note that the error bars shown in Fig.\ \ref{mass} reflect the Poisson uncertainties only; cosmic variance potentially contributes an additional $\sim$$13$\% uncertainty for populations of this number density in the volume sampled (cf.\ \citealt{tre08}).  The bar fractions and statistical uncertainties for each redshift, mass, morphology, SSFR, and environmental density bin measured here (and shown in Fig.\ \ref{mass}) are compiled in Table \ref{cosmosfracs} for reference.

At intermediate stellar mass the mean bar fraction of early-type disks ($f_\mathrm{bar} = 0.42\pm0.05$) is much higher than that of late-type disks ($f_\mathrm{bar}=0.19\pm0.03$) over the redshift interval explored.  However, at high stellar mass this trend is reversed with early-type disks less frequently barred ($f_\mathrm{bar}=0.20\pm0.04$) than late-type disks ($f_\mathrm{bar}=0.43\pm0.07$).  The bar fractions of galaxies with the lowest and highest specific star formation rates at both mass regimes closely follow the trends evident in the morphologically-selected early-type and late-type populations respectively---reflecting a close relationship between disk morphology and star formation rate at these epochs.  The bar fractions by large-scale environmental density reveal no relationship between the frequency of bars and environment at fixed mass.  However, we caution that the environmental densities from \citet{sco07} used here are based on photometric redshifts and averaged over $>$100 Mpc scales.  Hence, we are unable to confirm a null dependence of the bar fraction on density down to group scales with this dataset.  Finally, the total bar fraction is observed to build up by a factor of $\sim$2 between $z \sim 0.6$ and $z \sim 0.2$ at intermediate stellar mass, but at high stellar mass a null evolution scenario is favoured.

The local Universe bar fractions measured in a complementary sample of SDSS disks at $0.02 < z < 0.04$ described in the Appendix are also shown in Fig.\ \ref{mass} for comparison.  At intermediate stellar mass these low redshift datapoints confirm that the bar fraction in early-type disks remains constant, or even exhibits a slight decline, from $z \sim 0.2$ to $z \sim 0$ while assembly of the late-type barred disk population continues to the present day.  At high stellar masses these local Universe results are consistent with the null evolution of the early-type bar fraction since $z \sim 0.6$ evident in the COSMOS dataset.  At face value, the bar fraction in late-type disks at high mass in the SDSS at $0.02 < z < 0.04$ is much lower than that at $0.2 < z < 0.6$ measured in COSMOS; however, the statistical uncertainites are rather large due to the low number density of this galaxy population.  Indeed, there is a clear dependence of the total bar fraction on mass in both our low and high redshift samples, supporting our main conclusions.

\section{Discussion}\label{discussion}
Several observational studies indicate that the large/massive disk galaxy population has reached a mature state by redshift unity. These studies include the null evolution in the disk galaxy mass-size relation \citep{bar05}, the size function of large disks \citep{lil98,cam07,sar07}, the $B$-band Tully-Fisher relation for massive disks \citep{zie02}, and the morphological mix of massive galaxies \citep{oes09} since $z \sim 0.7$$-$1, as well as the relative paucity of highly disturbed and highly diffuse galaxies at $z \la 1$ \citep{con04}. Our work adds an important additional piece of information:  There exist substantial populations of barred early- and late-type disks---whether defined by morphology or SSFR---at $M > 10^{11}$ $M_\odot$ out to $z \sim 0.6$.  At such high stellar masses the fractions of barred disks in each of these galaxy classes remain roughly constant over the $\sim$3.3 Gyr that separate the $z \sim 0.6$ from the  $z \sim 0.2$ epochs.  Thus, already at the intermediate redshifts explored in our study, spectral and morphological transformations within the most massive disk galaxies have largely converged, to such a level of detail, to the familiar Hubble sequence that we observe in the local Universe.

In contrast, at intermediate stellar mass ($10^{10.5} < M < 10^{11}$ $M_\odot$) there is a substantial build-up in the bar fractions of early- and late-type disks, and of the total disk population (by a factor $\sim$2), over the past $\sim$3.3 Gyr of cosmic time.  At such intermediate mass scales, therefore, the disk galaxy population---and the disk-galaxy sector of the Hubble sequence---continues structural evolution until at least $z\sim0.2$.  The preference for (large-scale) bars in intermediate mass galaxies to reside in early-type disks---a trend also noted by \citet{she08} in their analysis of the COSMOS field---is perhaps a reflection of their earlier formation times compared to late-type disks of similar mass, i.e., these galaxies have evolved for longer in a dynamical state conducive to bar formation.

Interestingly, as the bar fraction reveals no dependence on large-scale environmental density at fixed mass, it would appear that any enhancement in the minor merger rate with environment (e.g.\ \citealt{hei09}) at these mass scales and redshifts does not lead to significant disk heating (cf.\ \citealt{hop08}), which could strongly impede bar formation \citep{ath86}.  Although, an increase in the rate of tidally-triggered bar formation in dense environments could also be masking this effect.  We stress, however, that the `environment' in our analysis, based on photometric redshifts, is averaged over $>$100 Mpc scales.  The study of the bar fraction as a function of the group environment at early epochs is an important open question that we are exploring using the zCOSMOS group catalog of \citet{kno09}.  Moreover, given the importance of the group environment in determining the properties of local galaxies (e.g.\ \citealt{rob10}), we will also soon be using the ZENS dataset (Carollo et al., in prep.) to construct a key benchmark for the barred galaxy fraction in local groups (Cameron et al., in prep.).

Finally, our study  also reveals the importance of investigating the evolution of barred galaxies at fixed stellar mass: this provides evidence for  a  substantial  change in the relative fractions of bars in early- and late-type disks, and in the evolution of the total  bar fraction, when stellar mass straddles across the `critical' mass scale of $\sim$$10^{11} M_\odot$.  The close relationship between galaxy evolutionary histories and stellar mass has also been demonstrated recently via studies of local galaxy morphologies \citep{bun05} and colours \citep{bal06}, as well as in high redshift samples \citep{kov09,bol09,tas09}.  We therefore highlight the importance of constructing even larger, mass-limited disk galaxy samples at low and high redshift for improving our understanding of bar formation---an endeavour that will be much assisted by various ground-based (UKIDSS, VIKING) and space-based (GOODS-ERS, HUDF09) near-infrared imaging surveys now under way, and their successors.

\cleardoublepage

\appendix
\section{Low Redshift Comparison Sample}
To provide a local Universe benchmark for comparison against the high redshift bar fractions quantified here in the COSMOS dataset we have constructed a complementary sample of low redshift disks drawn from the Sloan Digital Sky Survey (SDSS)\footnote{Whilst there already exists a valuable database of SDSS systems artificially redshifted to $0.7 < z < 1.2$ and matched to the PSF and signal-to-noise of the COSMOS ACS I-band imaging provided by \citet{kam07}, the effective resolution of these artificially redshifted galaxies is much lower than that for the $0.2 < z < 0.6$ COSMOS disk sample employed in this study, and thus not suitable for the present analysis.}.  From the visually-classified SDSS galaxy catalog presented by \citet{nai10} we have selected a subsample of 651 disk galaxies (S0$-$ to Sd, $-3 \le$ visual class $\le 7$) at $0.02 < z < 0.04$ with $M > 10^{10.5}$ $M_\odot$, axial ratio $> 0.7$ (i.e., disk inclination $\la 45$ deg), and $R_e > 2.2$ kpc.  At these redshifts the physical resolution scale (0.6-1.1 kpc/FWHM) and rest-frame wavelength coverage ($\sim$4000-7000 $\AA$) of the ground-based SDSS $g$ and $r$-band imaging are roughly comparable to those of galaxies in our primary sample at $0.2 < z < 0.6$ in the space-based COSMOS $I$-band imaging ($\sim$4300-7000 $\AA$ and 0.4-0.8 kpc/FWHM).

To ensure consistency with our selection of barred galaxies in COSMOS we have applied an identical ellipse-fitting bar detection procedure to the SDSS comparison sample.  For this purpose we downloaded the $u$, $g$, $r$, $i$, and $z$-band images of all 651 disk galaxies in our local Universe sample from the SDSS imaging server (\texttt{http://das.sdss.org/imaging/}).  Following the process described in Section \ref{detection}, we recovered elliptical isophote profiles for each galaxy in the $g$ and $r$-bands using \texttt{ellipse}, and identified (strong) barred galaxy candidates as those objects with maximum ellipticity $e \ge 0.4$ and a subsequent drop of $\Delta e \ge 0.1$.  Visual inspection of the $u$, $g$, $r$, $i$, and $z$-band images of the barred galaxy candidates was then used to construct master catalogs of confident $g$ and $r$-band bar detections.  During the visual inspection process 6 galaxies with axial ratio $> 0.7$ were noted to be bulge-dominated S0s with near edge-on (rather than face-on) disks and were thus excluded in the subsequent analysis.

\begin{table*}
\caption{Bar Fractions by Mass, Morphology, SSFR, and Environmental Density in an SDSS Local Universe Comparison Sample}
\label{sdssfracs}
\begin{tabular}{llrrr}
\hline \hline
Bar Catalog & Stellar Mass$^a$ & Morphological Type & Specific Star Formation Rate$^a$ & Environmental Density$^a$\\
\hline
$r$-band ellipse fit & Intermediate & Early-Type : $f_\mathrm{bar} = 0.40 \pm 0.03$ & Low SSFR :  $f_\mathrm{bar} = 0.49 \pm 0.04$ & High density : $f_\mathrm{bar} =  0.48 \pm 0.04$\\
$r$-band ellipse fit & Intermediate & Late-Type : $f_\mathrm{bar} = 0.41 \pm 0.03$ & High SSFR :  $f_\mathrm{bar} =  0.34 \pm 0.03$ & Low density : $f_\mathrm{bar} = 0.39 \pm 0.03$\\
$r$-band ellipse fit & High & Early-Type : $f_\mathrm{bar} = 0.13 \pm 0.06$ & Low SSFR :  $f_\mathrm{bar} =  0.21 \pm 0.09$ & High density : $f_\mathrm{bar} = 0.21 \pm 0.09$\\
$r$-band ellipse fit & High & Late-Type : $f_\mathrm{bar} = 0.22 \pm 0.08$ & High SSFR :  $f_\mathrm{bar} = 0.05 \pm 0.05$ & Low density : $f_\mathrm{bar} = 0.16 \pm 0.08$\\
\hline
\end{tabular}
\flushleft{$^a$ Here we define intermediate stellar mass as $10^{10.5} < M < 10^{11}$ $M_\odot$ and high stellar mass as $M > 10^{11}$ $M_\odot$, and we identify the low and high SSFR (or environmental density) populations as the lower and upper thirds of the SSFR (or environmental density) distribution in each mass and redshift bin.  The SDSS disk sample examined is also restricted to systems with axial ratio $> 0.7$ and $R_e > 2.2$ kpc as described in this Appendix.}
\end{table*}

\begin{figure}
\begin{center}
\includegraphics[width=8.4cm]{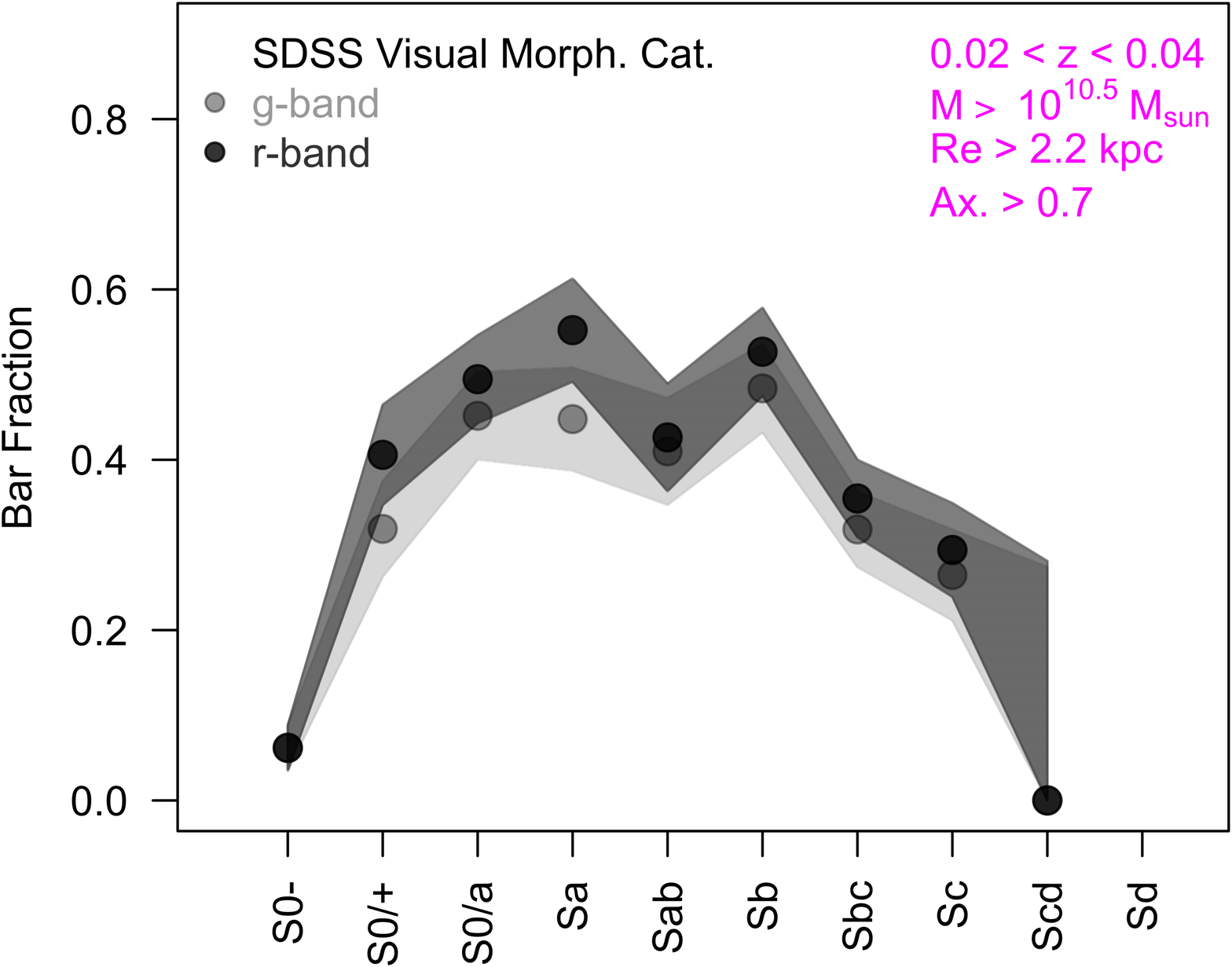}
\end{center}
\caption{The bar fraction by visual morphological class for massive ($M > 10^{10.5}$ $M_\odot$) disks at axial ratio $> 0.7$ and $R_e > 2.2$ kpc in the local Universe ($0.02 < z < 0.04$).  The visual morphological classifications are from the \citet{nai10} catalog.  The bar fractions recovered via application of the ellipse-fitting method (described in Section \ref{detection}) to the SDSS $g$- and $r$-band images of each galaxy are compared here to demonstrate the minimal dependence of bar detection efficiency on (optical band) wavelength across all morphological types.}
\label{sdssbars}
\end{figure}

From the 588 intermediate mass ($10^{10.5} < M < 10^{11}$ $M_\odot$) disk galaxies in our local Universe sample we identify 214 barred systems (a bar fraction of 36\%) in the $g$-band and 240 barred systems (41\%) in the $r$-band.  From the 57 high mass ($M > 10^{11}$ $M_\odot$) disk galaxies we identify only 9 barred systems (16\%) in the $g$-band and 10 barred systems (18\%) in the $r$-band.  The difference of $\sim$12\% between the bar fractions recovered here in the $g$ and $r$-bands is slightly higher than that of $\sim$5\% reported in the analysis of a high resolution ($D < 100$ Mpc) SDSS sample by \citet{she08}.  However, we note that the majority of bars missed by our $g$-band ellipse-fitting lie in compact galaxies with $2.2 < R_e \la 3$ kpc.  Hence, we suspect that the larger PSF FWHM in the SDSS $g$-band images (FWHM $\sim$ 1.5 arcsec) relative to the $r$-band (FWHM $\sim$ 1.4 arcsec) was the cause of the bar detection failures for these compact systems.  At $R_e > 3$ kpc the difference in recovered bar fractions between the $g$ and $r$-bands in our sample is only 6\%.  In Fig.\ \ref{sdssbars} we contrast the dependence of bar fraction on disk morphological type (at $R_e > 2.2$ kpc) recovered using our $g$ and $r$-band ellipse-fit catalogs in order to demonstrate the minimal dependence of bar detection efficiency on (optical band) wavelength across all morphological types.

In Table \ref{sdssfracs} we present the bar fractions in our local Universe comparison sample subdivided by mass, morphological type, SSFR, and environmental density as performed for our high redshift disk sample in Section \ref{results}.  Specifically, we employ the \citet{kau03} stellar masses to separate our SDSS sample into intermediate mass systems at $10^{10.5} < M < 10^{11}$ $M_\odot$ and high mass systems at $M > 10^{11}$ $M_\odot$.  Early-type and late-type disks are defined as those with visual morphological classes S0- to Sa and Sb to Sd, respectively.  To define our low (high) SSFR and density populations we adopt the \citet{bri04} SSFR estimates and \citet{bal06} environmental densities (average of $\Sigma_4$ and $\Sigma_5$), and identify the lowest (highest) thirds of each distribution.

A comparison of these local Universe bar fractions against those recovered at high redshift in COSMOS is presented in Section \ref{results}.  As described therein, the SDSS bar fractions by stellar mass and spectral/morphological type confirm the evolutionary trends observed for massive barred galaxies over the redshift interval $0.2 < z < 0.6$ in COSMOS.

\end{document}